--------------------(LATEX FILE)--------------------------
\documentstyle{article}
\begin{document}
\twocolumn[
\begin{center}
\large
TUNNELING IN PAIRED FRACTIONAL QUANTUM HALL STATES:
CONDUCTANCE AND ANDREEV REFLECTION OF NON-ABELIONS

\vspace{0.3cm}
\normalsize
Ken-ichiro Imura\\

\small
Department of Applied Physics, University of Tokyo,
Hongo, Tokyo 113-8656, Japan\\

\vspace{0.3cm}
\normalsize
Kazusumi Ino\\

\small
Institute for Solid State Physics,
University of Tokyo, Roppongi, Tokyo 106, Japan\\
\end{center}

\vspace{0.3cm}
\small
We study the edge transport properties of paired fractional
quantum Hall (FQH) states---
the Haldane-Rezayi (HR), Moore-Read (Pfaffian)
and Halperin (331) states.
A table of exponents is given for the tunneling between the
edges of paired FQH states in gated 2D structures and the tunneling
into the edge of FQH states from a normal Fermi liquid (N).
It is found that HR, Pfaffian and 331 states have
different exponents for quasiparticle tunneling.
For the tunneling through a FQH-N junction,
we propose unusual Andreev reflection processes that may also
probe the non-abelian FQH states.

\begin{center}
\begin{tabular}{lcccccccc} \hline\hline
Edge state&
$\Psi_{elec}$&
$\Psi_{qp}$&
$\Delta_{elec}$& 
$\Delta_{qp}$&
$I-V$&
QPT&
DIGA&
ET\\ \hline
331 (boson) &
$e^{i\pm \phi_s}e^{i2\phi_c}$&
$e^{i\phi_{\uparrow,\downarrow}}$&
3/2&
3/16&
$I \propto V^3$&
$\Delta G\propto T^{-5/4}$&
($G\propto T^6$)&
$G\propto T^4$
\\
331 (Dirac)&
$\Psi_D^\alpha e^{i 2\phi_c}$&
$\mu e^{i\phi_c/2}$&
3/2&
3/16&
$I \propto V^3$&
$\Delta G\propto T^{-5/4}$&
($G\propto T^6$)&
$G\propto T^4$
\\
Pfaffian&
$\Psi_M e^{i 2\phi_c}$&
$\sigma e^{i\phi_c/2}$&
3/2&
1/8&
$I \propto V^3$&
$\Delta G\propto T^{-7/4}$&
&
$G\propto T^4$
\\
HR ($c=-2$ CFT)&
$\partial\theta^\alpha e^{i2\phi_c}$&
$\eta e^{i\phi_c/2}$&
2&
$-1/16$&
$I \propto V^4$&
$\Delta G\propto T^{-9/4}$&
&
$G\propto T^6$
\\
HR (Gurarie-Flohr-Nayak)&
\multicolumn{8}{c}
{Same as the Dirac fermion description of the 331 state}
\\
HR (Lee-Wen)&
&
&
2&
1/8&
$I \propto V^4$&
$\Delta G\propto T^{-7/4}$&
&
$G\propto T^6$
\\ \hline
\multicolumn{5}{l}
{$\nu =1/2$ compressible state}&
$I \propto V^3$&
\multicolumn{3}{c}{}
\\
\multicolumn{5}{l}
{Experiment (A.M. Chang)}&
$I\propto V^2$&
\multicolumn{3}{c}{}
\\ \hline\hline
\end{tabular}
\end{center}

\noindent
Table I:
Summary of the results. The electron and quasiparticle
annihilation operators, their scaling dimensions, $I-V$
characteristics of the tunneling through a FQHL-N junction
and the conductance $G(T)$ through a PC is assembled.
The exponents which have appeared in Ref. \cite{GFN} and \cite{LW}
are also shown.
$G(T)$ for the QPT model is given by
$G(T)=e^2/(2h) - \Delta G$.
DIGA gives $G\propto T^6$ when the spin-flip scattering at the
point contact is possible,
whereas $G\propto T^4$ in the absence of spin-flip
scattering.

\vspace{0.3cm}
]

\noindent
The physical properties
of the edge of incompressible fractional quantum Hall
liquids (FQHL) have received a great deal of attention in recent years
\cite{PQHE},
which can be probed by the tunneling experiments.
They include the tunneling between the edges of FQHL 
in gated 2D structures \cite{T4},
and the tunneling into the edge of FQHL from a 3D
normal Fermi liquid \cite{V3}.
These experiments have been performed for $\nu=1/3$ Laughlin state
\cite{T4,V3}, and
support the idea that the edge mode of FQH state
is described as a chiral Tomonaga-Luttinger (TL) liquid \cite{Wen1}, 
showing the power-law dependence of the conductance on the
temperature and voltage \cite{1/3}.
The tunneling experiment has been extended to other filling
factors such as $\nu=1/2$ \cite{V2}, where the 2D electron system may
be compressible \cite{HLR}. 
Shytov, Levitov and Halperin \cite{SLH} have studied the tunneling into
the edge of a compressible state, and found a non-Ohmic
tunneling conductance of $I \propto V^3$ at $\nu=1/2$,
while the experiment have shown another exponent of $I \propto V^2$
\cite{V2}.

FQH plateaus are observed in $\nu= 5/2$ single layer
and $\nu= 1/2$ double layer systems \cite{exp}
in addition to the usual odd denominator filling factors.
Three possible states have been proposed to explain
such incompressible plateaus at even denominator filling
factors.
They are the Haldane-Rezayi (HR) \cite{HR}, Moore-Read (Pfaffian)
\cite{MoR,GWW} and Halperin (331) \cite{Helv} states.
These states are regarded as the FQH analogues of the BCS state.
The pairing symmetry is $p$-wave with $S_z=1, 0$ for the Pfaffian
and 331 states, respectively, and $d$-wave for the HR state
\cite{Ho}.
The HR and Pfaffian states are of theoretical interest
due to some novel features in their topological
ordering such as non-abelian statistics of quasiparticles
\cite{MoR}.
On the other hand, the 331 state belongs to the standard
abelian hierarchy construction\cite{Gir},
but it can also be interpreted as a paired state
\cite{Helv}.
Paired FQH states at $\nu = 1/2$ have also attracted considerable
attentions in regard to the recently observed
anomalous MIT (metal-insulator transition) in 2D \cite{Krav}. 

It has been observed \cite{MoR,WW} that
the bulk wave functions of a number of FQHL's could be interpreted  
as the conformal blocks of certain 
conformal field theories (CFT).
In the case of Pfaffian, 331 and HR states,
the corresponding CFT's have
the central\\
charges $c=1/2+1$, $c=1+1$ and $c=-2+1$, respectively.
This point of view is of great power
because the edge states of FQHL are described by the same
CFT as the bulk wave functions,
which is related to
the equivalence \cite{Wit} between the states of a  
Chern-Simons theory and the conformal blocks in an
associated CFT.
However this general statement is valid only for
the unitary CFT ($c>0$), and $c=-2$ CFT is not the case.
In the latter case, a more direct approach is necessary.
The states of edge CFT can be enumerated by a direct
construction of the corresponding
lowest Landau level wave functions which are the exact
zero-energy eigenstates of certain model Hamiltonian
\cite{GWW}.
Generalizing this construction,
Milovanovic and Read \cite{MiR}
have studied
the edge modes of the Pfaffian, 331 and HR states,
and found that they have
fermionic excitations in addition to
the usual bosonic charge fluctuation excitations.
The fermionic fields in the Pfaffian, 331 and HR cases are
Majorana ($c=1/2$), Dirac ($c=1$)
and scalar ($c=-2$) fermions, respectively.

It should be noted that
there is a close relationship between the
$c=-2$ and $c=1$ CFT's \cite{GL,GFN}.
In terms of the mode expansion, the Dirac fermion has
half-integral momenta in the untwisted sector and
integral momenta in the twisted sector,
while the opposite is true for the scalar fermion in $c=-2$.
We can map the twisted sector of $c=-2$ CFT into the
untwisted one of $c=1$ \cite{GL}.
It should also be noted that
the $c=-2$ CFT,
has some difficulties such as the negative
scaling dimension of the quasiparticle propagator \cite{LW}. 
Then Gurarie, Flohr and Nayak \cite{GFN},
inspired by the analogy between thw two CFT's,
proposed that the edge mode of the HR state
is described by a chiral Dirac fermion
($c=1$ gaussian field at a compactification radius $R=1$).

In this paper,
we study the tunneling into and between the paired FQH
states.
Low-energy transport properties are determined by the most
relevant operators that can contribute to the tunneling.
Hence we first identify that operator
in each of the experimental geometries.
Then it is straightforward to study the power law behaviors
of the tunneling conductances.
For the HR state we consider three possibile cases.
One is that the naive $c=-2$ CFT (conformal field theory) works.
Another possibility has been proposed by Gurarie, Flohr
and Nayak, who described the edge mode of the HR state
as the Dirac fermion in addition to the usual charge boson,
which is also the case with the 331 state.
The other scenario is due to Lee and Wen \cite{LW},
who calculated the exponents of quasiparticle propagator
without using CFT.
We have studied the scaling behaviors of the tunneling conductance
for these possible scenarios,
and the results are summarized in Table I.
The details of the derivation are given below.
For the tunneling through a FQHL-N junction, we propose
the Andreev reflection of non-abelions\cite{MoR}, 
which is a non-abelian analogue of the Andreev reflection
of Laughlin quasiparticles \cite{SCF}.

We begin with the boson description of the 331 state.
The effective Lagrangian density reads \cite{Wen2}
\begin{equation}
{\cal L} =
{1 \over 4\pi}
\partial_x \phi^\alpha
(
K_{\alpha\beta}
\partial_t \phi^\beta
-
V_{\alpha\beta}
\partial_x \phi^\beta),
\end{equation}
where
\begin{equation}
K = \left(
\begin{array}{rr}
3 & 1 \\
1 & 3
\end{array}
\right),
\ \ \
V = \left(
\begin{array}{cc}
v_\uparrow & v_{int}\\
v_{int} & v_\downarrow
\end{array}
\right).
\end{equation}
and $\alpha,\beta = \uparrow, \downarrow$
are spin indices.
Repeated indices are summed over.
The boson fields
$\phi^{\uparrow, \downarrow}$ corresponding to up and down spins
are related to the charge and spin bosons as
$\phi^c=\phi^\uparrow + \phi^\downarrow$, 
$\phi^s=\phi^\uparrow - \phi^\downarrow$.
The matrix $V$ depends on the interaction in general, i.e.,
on the details of edge structure.
Note that the charge and spin modes propagate in the same
direction corresponding to the two positive eigenvalues of
matrix $K$.
In this case,
the scaling behavior of the tunneling through
a point contact becomes universal
after integrating out the unimportant degrees of freedom
\cite{2/3}.

In the case of electron annihilation, the
most relevant operators are
$\Psi_{elec}^\alpha\sim
e^{\pm i\phi^s}e^{i2\phi^c}
=e^{i(3\phi^\uparrow + \phi^\downarrow)}$,  
$e^{i(\phi^\uparrow + 3\phi^\downarrow)}$.
Note that in our notation,
the scaling dimensions $\Delta$ of the vertex operator
$e^{i q \phi}$ is $\Delta = q^2 K$, where $\sqrt{2K}=1/R$.
In the present case
$K=1/2$ for $\phi^s$, $K=1/4$ for $\phi^c$,
and for the simple Laughlin case, $K=\nu$.
Now we can write down the scaling dimensions $\Delta_{elec}$ of
$\Psi_{elec}$ as
$\Delta_{elec}=1/2+2^2(1/4)=3/2$.
The scaling dimension $\Delta_{elec}$
is related to the exponent $g_{elec}$ of electron propagator
at $x=0$,
$\langle\Psi_{elec}(t)$
$\Psi_{elec}^\dagger (0)\rangle
\sim 1/t^{g_{elec}}$,
as $g_{elec}=2\Delta_{elec}$.

In the point contact geometry,
quasiparticles can tunnel between the edges
in the weak scattering regime.
The most relevant quasiparticle operators are
$\Psi_{qp}^\alpha\sim 
e^{i\phi^\alpha}=
e^{\pm i\phi^s/2}
e^{i\phi^c/2}$,
which has the scaling dimension
$\Delta_{qp}=(1/2)^2(1/2)+
(1/2)^2(1/4)=3/16$.
The quasiparticle propagator obeys a power law
$\langle\Psi_{qp}(t)\Psi_{qp}^\dagger (0)\rangle$
$\sim 1/t^{g_{qp}}$,
where $g_{qp}=2\Delta_{qp}$.  

The next step is to recover all the above results
in terms of the Dirac fermion theory ($c=1, R=1$).
The electron operators can be written as
$\Psi_{elec}^\alpha\sim\Psi_D^\alpha e^{i2\phi^c}$,
where $\Psi_D^\alpha$ is a Dirac fermion ($R=1$ i.e. $K=1/2$)
with spin $\alpha$.
While the interpretation of the quasiparticle operators is
less trivial, since they are half-flux quantum quasiparticle.
They gain an AB phase $\pi$ from the non-abelian spin part
in addition to the usual $\pi$ phase from the abelian charge sector,
as they go around an electron (and not a pair of electrons!) 
The former is related to the OPE (operator product expansion)
\cite{MoR,WW},
\begin{equation}
\Psi_D^\alpha (z)\mu(w)\sim{1\over (z-w)^{1/2}}
\left(
\tilde{\mu}^\alpha(w)+O(z-w)
\right),
\end{equation}
where
$\mu$ is a Dirac theory twist field with the conformal weight
$\Delta =1/8$ \cite{GL}, 
and $\tilde{\mu}^\alpha$ is a disorder field.
Including both the abelian and non-abelian parts,
the single-valuedness of the wave functions is guaranteed.
Now we can write down 
the quasiparticle operator as
$\Psi_{qp}\sim
\mu e^{i \phi_c/2}$,
which recovers the scaling dimension
$\Delta_{qp}=1/8+1/16=3/16$.

We can follow the same line of arguments
for the Pfaffian state, where the Dirac fermion $\Psi_D$ is
replaced by a Majorana fermion $\Psi_M$ and
the twist operator $\mu$ by an Ising spin $\sigma$ \cite{MoR},
$\Psi_{elec}\sim
\Psi_M e^{i 2\phi^c}$,
$\Psi_{qp}$
$\sim\sigma e^{i \phi^c/2}$.
The only difference is that the Ising spin
$\sigma$ has the conformal weight $\Delta =1/16$
instead of 1/8,
whereas the Majorana fermion
$\Psi_M$ has the same conformal weight $\Delta =1/2$,
as the Dirac fermion.
Hence
$\Psi_{elec}$ have the same scaling dimensions
$\Delta_{elec}=3/16$ as the 331 state,
whereas $\Delta_{qp}=1/16+1/16=1/8$.

Let us advance to the $c=-2$ CFT case,
the theory of scalar fermion $\theta^\alpha$,
which has an unusual scaling dimension $\Delta =0$.
The electron operator can be written in terms of
$\theta^\alpha$ as 
$\Psi_{elec}^\alpha
\sim\partial \theta^\alpha e^{i2\phi^c}$,
the scaling dimension of which is
$\Delta_{elec}=1+0+2^2(1/4)=2$ \cite{MiR}.
There is also a twist operator $\eta$ in the $c=-2$ CFT,
which obeys to the OPE similar to (3), 
and has the scaling dimension $\Delta =-1/8$ \cite{WW}. 
The most relevant quasiparticle operator can
be written in terms of $\eta$ as 
$\Psi_{qp}\sim\eta e^{i\phi^c/2}$,
which has a negative scaling dimension \cite{LW}
$\Delta_{qp}=-1/8+1/16=-1/16$.
Later we will comment on the validity of the naive $c=-2$ CFT
description.

\vspace{0.2cm}
\noindent
{\it Tunneling into the incompressible states}---
Let us consider a FQHL-N (normal metal) junction.
The edge modes of FQHL are coupled to the Fermi liquid lead.
The tunneling through the junction is described by
the operators
$\mbox{(FQHL-N tunneling)} \sim \Psi_{elec}\Psi_{lead}^{\dagger}$,\\ 
which is the FQHL-analogue of Andreev reflection
\cite{SCF}.
The tunneling current through the junction scales as \cite{1/3}
\begin{equation}
I \sim {\mbox{(FQHL-N tunneling)} \over \Lambda}
\sim \Lambda^{g_{elec}+1-1}
\end{equation}
with $\Lambda$ being a high frequency cutoff.
Hence the current-voltage (I-V) characteristics are obtained as 
\begin{equation}
I \propto
\left\{
\begin{array}{ll}
V^3 & (\mbox{331, \ Pfaffian})\\
V^4 & (c=-2 \ \mbox{CFT})
\end{array}.
\right.
\end{equation}
for $V \gg T$. For $V \ll T$ we have the Ohmic behavior
$I\propto V$.
The observed exponent ($I \propto V^2$)
is smaller than the theoretical predictions.
In the above analysis, we consider the point-like contact
between the FQHL and Fermi liquid,
assuming that the tunneling is not uniform along the edge
and that at very low temperatures only
the tunnling at the most dominant position
becomes important.
Tunneling conductance $G$ in the strong coupling
can be calculated using the weak-strong coupling duality of the TL
liquid \cite{CF}.
Since the abelian charge sector completely determines the charge
conductance, $G$ saturates to $(2/3)e^2/h$.
However the observed two terminal conductance $G_2$ of the
device may be different due to the various possibilities of
contacts between the reservoir and the FQH state \cite{CF}.

Next we consider the Andreev reflection of non-abelions.
In the Andreev reflection in normal-metal/
superconductor (N-S) junction,
an incident electron from the N side is back-reflected
as a hole while transferring charge $-2e$ to the S side.
In the FQHL-N tunneling, FQHL plays the role of normal
metal (N), whereas the electron gas reservoir (N) plays the role of
superconductor (S) \cite{SCF}.
Since the abelian (charge) case has been studied in Ref. \cite{SCF},
here we concentrate on the non-abelian (spin) cases.
In Fig.~1, we show some Andreev processes of non-abelions
which are determined by the fusion rules of CFT \cite{MoR,GFN}:
$\sigma\times\Psi_M=\sigma$
for $c=1/2$, and
$\eta\times\partial\theta^\alpha=\tilde{\eta}^\alpha$
for $c=-2$, where $\tilde{\eta}^\alpha$ is a dimension $3/8$
operator. 
For the HR ($c=-2$ CFT) case, pair creations of spin may
occur, since the incoming quasiparticle $\eta e^{i\phi^c/2}$
is spinless, 
whereas the transmitted electron
$\Psi_D^\uparrow e^{i2\phi^c}$ has a spin.
A quantitative discussion will be given elsewhere.

\vspace{0.3cm} 
\noindent
{\it Tunneling through a constricted point contact}---
Consider a two-terminal Hall bar geometry
where the bulk FQH liquid has both top and bottom edges.
Applying the negative gate voltage to squeeze the Hall bar,
one can introduce the depleted region of electrons.
This structure, called a point contact,
introduces the backward scattering between the edges due to the
quasiparticle tunnneling (QPT) through the bulk FQH liquid,
$QPT \sim \Psi_{qp}\Psi_{qp}^{\dagger}$.
Another model is the electron tunneling (ET) model, where the
depleted region is considered to be a vacuum, and the
electron can tunnel through this region between the
left and right FQH liquids,
$ET \sim \Psi_{elec}\Psi_{elec}^{\dagger}$.
The tunneling current $I_{PC}$ through a PC scales as,
$I_{PC} \sim ET/\Lambda
\sim \Lambda^{2g_{elec}-1}$.
Hence the tunneling conductance $G(T)$ is given by \cite{1/3,Wen2}
\begin{equation}
G(T)
\left\{
\begin{array}{l}
={1 \over 2}{e^2 \over h}-cst.\times T^{2g_{qp}-2}\ \ \
(\mbox{QPT model})\\
\propto T^{2g_{elec}-2}\ \ \ (\mbox{ET model})
\end{array},
\right.
\end{equation}
with the crossover temperature $T_{QPT}$ being
\begin{equation}
T_{QPT}=\left(
(QPT)_0 \over \Lambda_0^{g_{qp}}
\right)
^{1/(1-g_{qp})}.
\end{equation}
where $(QPT)_0$ and $\Lambda_0$
are the bare strength of QPT and the bare cutoff of the order
of the band width, respectively.
$T_{QPT}$ is the temperature where the perturbation expansion breaks
down. The crossover at intermediate couplings is non-perturbative
but it may be accessible through the Bethe ansatz \cite{BA}.
In terms of the boson description,
the strong scattering regime ($T \ll T_{QPT}$) of the QPT model 
can be studied using the
dilute instanton gas approximation (DIGA) in
the dual QPT model \cite{DIGA}.
Note that DIGA is not applicable to the other fermionic
descriptions.

For the 331 state, 
in the absence of spin-flip scattering at the point contact,
DIGA and the ET model give the same exponents,
$G(T)\propto T^4$.
Introducing the spin-flip scattering at the point contact,
DIGA gives another exponent,
$G(T)\propto T^{6}$.
The exponents are robust against the details of edge structure,
reflecting the same chirality of two edge modes.
Substituting the obtained $g_{qp}$ and $g_{elec}$
to the formula (6),
we can write down the conductance $G(T)$
for the Pfaffian and the HR states.
The results are summarized in Table I.
The exponents which have appeared in Ref. \cite{GFN} and \cite{LW}
are also shown.

\vspace{0.3cm}
\noindent
{\it Discussion and Conclusions}---
We have studied the edge transport properties of paired FQH states---
331, Pfaffian and HR states,
which have fermionic edge excitations,
in addition to the usual charge bosons,
described by Dirac, Majorana and scalar fermions,
respectively.
It is found that they have different exponents for
the QPT model in FQH-FQH tunneling.
However for the $c=-2$ CFT,
it has been pointed out \cite{GFN,LW} that
it has some difficulties such as the negative
scaling dimension of the quasiparticle propagator. 
To remedy them,
Gurarie, Flohr and Nayak \cite{GFN}
proposed the idea that the edge mode of the HR state
is described by the $c=1, R=1$ CFT.
While Lee and Wen \cite{LW}
calculated the electron and quasiparticle exponents
of the HR state to show that 
$g_{elec}$ remains 3, whereas
$g_{qp}$ should be replaced by the Pfaffian's value 1/4.

In conclusion, the table of exponents is given for the tunneling
into and between the paired FQH states.
We propose the Andreev reflection of non-abelions expected
from the fusion rules of CFT.
The non-abelian states may also be probed 
by observing the persistent edge current coupled to an
Aharanov-Bohm flux \cite{AB}.
We hope that the edge mode of FQHL
will serve as an experimental probe, i.e., as a ``window'' 
to identify the bulk topological order.

\vspace{0.3cm}
{\it Acknowledgements}---
The authors would like to thank N. Nagaosa,
M. Yamanaka and S. Murakami
for stimulating discussions.
This work was supported by COE and Priority
Areas Grants from the Ministry of Education, Science
and Culture of Japan.
One of the authors (Ken. I.) is also supported by JSPS
Research Fellowships for Young Scientists.

\end{document}